# Induced Representations of Tensors and Spinors of any Rank in the SHP Theory

Lawrence P. Horwitz

Raymond and Beverly Sackler Faculty of Exact Science, School of Physic, Tel Aviv University, Ramat Aviv 69978, Israel.

Physics Department, Bar Ilan University, Ramat Gan 52900, Israel.

Israel Physics Department, Ariel University, Ariel 40700, Israel

Email: larry@post.tau.ac.il

Meir Zeilig-Hess

Raymond and Beverly Sackler Faculty of Exact Science, School of Physic, Tel Aviv University, Ramat Aviv 69978, Israel.

Email: meirzh10@gmail.com

**Abstract**

We show that a modification of Wigner's induced representation for the description of a relativistic particle with spin can be used to construct spinors and tensors of arbitrary rank, with invariant decomposition over angular momentum. In particular, scalar and vector fields, as well as the representations of their transformations, are constructed. The method that is developed here admits the construction of wave packets and states of a many body relativistic system with definite total angular momentum. Furthermore, a Pauli-Lubanski operator is constructed on the orbit of the induced representation which provides a Casimir operator for the Poincaré group and which contains the physical intrinsic angular momentum of the particle covariantly.

# 1 Introduction

## 1.1 Foundations of the SHP theory

The theory of Stueckelberg-Horwitz-Piron (called SHP henceforth) was initiated by Stueckelberg's suggestion, according to which pairs of particles can be created and/or annihilated throughout the motion of a single particle, in a way which can be described by classical equations of motions in space-time[1]. The evolution parameter, $\tau$, of such particle trajectories is an invariant under Lorentz transformations. The metric, $\eta$, is taken henceforth to be the Minkowski flat metric with signature $(-1, +1, +1, +1)$.

Horwitz and Piron generalized Stueckelberg's idea in order to describe a system of particles with no internal degrees of freedom[2]. They obtained the relativistic generalization of the Hamilton equations of motion for a system of $N$ particles:

$$\begin{cases} \frac{dx_i^\mu}{d\tau} = +\frac{\partial K}{\partial p_\mu^i} \\ \frac{dp_i^\mu}{d\tau} = -\frac{\partial K}{\partial x_\mu^i} \end{cases} (1.1)$$

These equations are derived from a covariant Hamiltonian, $K(x,p)$, which is assumed to be a scalar function (under Lorentz transformations) of the phase-space variables $(x,p) = \{x_i^\mu, p_i^\mu\}, i = 1, \dots, N$.

The quantization of the theory is done by making the following postulates[1]:

• Space-time and energy-momentum variables become Hermitian operators which obey the following commutation relation:

$$[x_i^\mu, p_j^\nu] = i\hbar \eta^{\mu\nu} \delta_{ij} \ (1.2)$$

• The state of a particle is described by a vector, $|\psi_\tau\rangle$ which is an element of a Hilbert space, $\mathcal{H}$. Such a state is represented in the coordinate basis as: $\psi_\tau \in \mathbb{L}^2(\mathbb{R}^4; d^4x)$.

• The evolution of such a state in $\tau$ is unitary, in order to conserve in $\tau$ the inner product of any two quantum states, $|\psi_\tau\rangle, |\phi_\tau\rangle \in \mathcal{H}$:

$$\langle \phi_\tau | \psi_\tau \rangle = \int \phi_\tau^*(x) \psi_\tau(x) \, d^4x \ (1.3)$$

In particular, the normalization $\langle \psi_\tau | \psi_\tau \rangle = 1$, which is needed for a probabilistic interpretation of the theory, is then constant in $\tau$. The norm $|\psi_\tau(x^\mu)|^2$, is the probability density per unit space-time volume for an event $x^\mu$ to occur at instant $\tau$.

The unitary evolution manifests itself in an evolution equation, which generalizes that of Schrödinger for the relativistic regime:

$$i\hbar \frac{\partial |\psi_\tau\rangle}{\partial \tau} = K |\psi_\tau\rangle \ (1.4)$$

This equation is known as the Stueckelberg-Shrödinger equation. The covariant Hamiltonian, $K$, is the generator of the evolution in $\tau$. It is a Hermitian operator with respect to the inner product (1.3) and, if invariant, it commutes with the generators of the Lorentz group.

The evolution parameter, $\tau$, is the generalization of the Newtonian time parameter in the Stueckelberg-Schrödinger equation[3].

The introduction of $\tau$ allows for variety of quantum phenomena to occur, such as interference in both space and time[4], in accordance with the attosecond experiment of Lindner et. al.[5,6], as well as relativistically covariant formulations of diffraction[4], scattering[7] and the two-body problem[8].

It is assumed in the SHP framework[2] that a free spinless particle evolves dynamically according to the covariant Hamiltonian: $K_{free} = \eta_{\mu\nu} \frac{p^\mu p^\nu}{2M}$, where $M$ is a mass parameter. Based on the first two postulates, the momentum operator is quantized in the coordinate basis according to: $p_\mu = \frac{\hbar}{i} \partial_\mu \leftrightarrow \frac{\hbar}{i}\left(\frac{1}{c}\frac{\partial}{\partial t}, \frac{\partial}{\partial x}, \frac{\partial}{\partial y}, \frac{\partial}{\partial z}\right)$.

The commutation relations (1.2) give rise to the generalized uncertainty principle: $\Delta x_\mu \Delta p_\mu \geq \frac{\hbar}{2}$, for each Lorentz index, $\mu$ (no summation). In the case of spacial indices, the well known Heisenberg inequality is recovered; one finds rigorously the time energy uncertainty relation as well[9].

## 1.2 Wigner Rotations

The formulation of spin in a covariant way was first made by Wigner in his famous work[10]. The main result of Wigner's analysis to be used in our work is that a subgroup of rotations may be induced by the successive application of three Lorentz transformations. Thus, according to Wigner's method, if we apply a boost $L(p)$, succeeded by a general Lorentz transformation $\Lambda$, succeeded by the boost $L^{-1}(\Lambda p)$, we obtain (in terms of the fundamental representation of $SL(2, \mathbb{C})$, the covering group of O(3,1)) what is now called a Wigner rotation:

$$R(\Lambda, p) = L^{-1}(\Lambda p)\Lambda L(p) \quad (1.5)$$

In Eq. (1.5), the boost $L(p)$ is defined as the Lorentz transformation that takes the momentum from its rest frame value $p_0 = (mc, 0,0,0)$ ($m$ is the particle's mass) to a general timelike $p$. Therefore, the little group of rotations is parameterized by the momentum, with $p$ the stability vector of the little group.

The induced representations of the Lorentz group were studied by Mackey. He generalized the theory for any connected topological group which can be written as a semi-direct product of two of its subgroups, one of which is normal. The prescription for finding all irreducible representations of such groups, using the induced little groups, is now known as the Mackey-Frobenius theorem[11].

In the Hilbert space of quantum states, Lorentz transformations are realized by unitary operators. These operators are infinite dimensional as a result of the noncompactness of the Lorentz group manifold, $\ell_+^\uparrow$ [12]. However, the realization of the induced rotation given by Eq. (1.5) is finite dimensional and unitary, as the $SU(2)$ group manifold is compact.

The unitary operators of the induced rotations of Eq. (1.5) depend on momentum. This dependence in the transformation law of the wave-function destroys the covariance of the expectation value of the position operator. Furthermore, it is impossible in this framework to form wave packets of definite spin by integrating over the momentum variable, since this would add functions over different parts of the orbit with a different $SU(2)$ at each point. These problems are solved if we introduce a time-like vector, $n^\mu$, that stabilizes the induced representation of Eq. (1.5) instead of the 4-momentum[13].

As remarked in ref. [14], the vector $n^\mu$ may be associated with a timelike normal to the spacelike surfaces of Schwinger[15] and Tomonaga[16], since the representation achieved for the relativistic spin in this way generates an SU(2) on the spinor fields corresponding to rotation in the spacelike surface orthogonal to $n^\mu$. We shall, furthermore, construct in Section 4, the Pauli-Lubanski operator which provides a Casimir operator for the Poincaré group on the orbit at the point $n^\mu$ that corresponds to the intrinsic angular momentum of the particle in covariant form[1].

With the introduction of this timelike vector, the Lorentz group is generated by Eq. (1.5), and the states belong to the Hilbert space $\mathbb{L}^2(\mathbb{R}^{3,1}, d^4x; \mathbb{R}^3, d^3n) \times \Sigma$, where $\Sigma$ is the Hilbert-space of internal degrees-of-freedom. These states will be denoted in Dirac notation as the kets: $|n, \sigma\rangle$. This implies the fact that the particles have a definite stability vector $n$, a definite total spin (suppressed in this notation), and a definite spin projection $\sigma$ (on an arbitrarily chosen quantization axis).

When a general Lorentz transformation, $\Lambda \in \ell_+^\uparrow$, represented by a unitary operator $U(\Lambda)$ in the Hilbert-space, acts on a state of a particle, one can easily check, by the same method as used in arriving at (1.5), that:

$$U(\Lambda)|n, \sigma\rangle = U[L(\Lambda n)]|n_0, \sigma'\rangle D_{\sigma'\sigma}(\Lambda, n) \quad (1.6)$$

where $D(\Lambda, n)$ is the realization of the Wigner rotation. This rotation operator is given by the chain of operations that leaves the "rest frame" vector, $n_0$, invariant:

$$D(\Lambda, n) = U^{-1}[L(\Lambda n)]U(\Lambda)U[L(n)] \quad (1.7)$$

It was recently argued[14] that the elementary proof of the spin-statistics theorem, involving a $\pi$ rotation as equivalent to an exchange of two identical particles, can be carried through only if all identical particles are on the same point n of the orbit of the induced representation, and thus have the same $SU(2)$. Therefore, this modification of Wigner's analysis applicable to the SHP theory admits the definition of total angular momentum for a relativistic many body system, using $SU(2)$ Clebsch-Gordan coefficients for the addition of angular momenta. Furthermore, the fact that the little group representations do not depend on momentum admits the construction of wave-packets in the relativistic regime.

In this paper we show that the induced representations of the little group can be used to construct tensors and spinors of arbitrary rank with invariant decomposition into definite spins. In section 2 we obtain the covariant induced representations of definite spins. In 2.1 we point out the explicit steps that are to be performed in order to obtain representations of the little group for rank-1 tensors. In section 2.2 we generalize the procedure for arbitrary higher representations of the little group. Section 3 deals with the explicit construction of the fields of definite spin. In 3.1 we review the construction of the fundamental representation of spin 1/2. In 3.2 we construct scalar and vector fields and explore some possible dynamical properties related to them. In 3.3 we outline the method of construction of tensors and spinors of higher spin values. In Section 4 we construct a Pauli-Lubanski vector that contains the physical angular momentum of the particle in a covariant form and provides a second Casimir operator for the Poincaré group on the orbit of the induced representation. Conclusions and a discussion of the results are given in section 5.

---

[1] As Kaku [16] points out, the usual Pauli-Lubanski operator is only related to the angular momentum of the particle in the rest frame.

# 2 Induced Representations in the SHP Theory

In the previous section we presented the fundamental representation of the induced little group of $\ell_+^\uparrow$ for off-shell massive particles. The rotations that belong to the little group are obtained by a successive application of three Lorentz transformations. These transformations may be realized by matrices of finite dimension (which is 2 in the fundamental representation) belonging to the $SL(2,\mathbb{C})$ covering group of $\ell_+^\uparrow$. Since the manifold of the Lorentz group is non-compact, the $SL(2,\mathbb{C})$ realizations are not unitary. However, the successive application in Eq. (1.5) can be realized as a matrix that is both unitary and of finite dimension[12,18].

We now set forth to investigate the representations of the little group that are higher than the fundamental one.

## 2.1 The Next-to-Fundamental Representation

In order to investigate the induced representations of the Lorentz group for the case of vector fields, we need to look for the next representation of the little group. Higher representations of $SU(2)$ may be obtained, using the same technique, from the fundamental one by the use of Clebsch-Gordan coefficients[19]. Using the D matrices of Eq. (1.7) for the fundamental representation, a construction of the next-to-fundamental representations may be achieved by direct-multiplication of the elementary ones. These representations are obtained as follows:

1. We replace each of the $SL(2,\mathbb{C})$ factors in the chain operations of Eq. (1.5), by a direct product of two $SL(2,\mathbb{C})_a$ matrices ($a = 1,2$), each belonging to a different spin space, resulting in elements of the $SL(2,\mathbb{C})_1 \times SL(2,\mathbb{C})_2$ group. Since the manifold of $SL(2,\mathbb{C})$ is non-compact so is the manifold of the direct product group, hence each $SL(2,\mathbb{C})_1 \times SL(2,\mathbb{C})_2$ element cannot be both finite dimensional and unitary.
2. We first perform the simple multiplications between the matrices in each spin-space independently and only then carry out the outer product of the two resultant matrices. In each spin-space we obtain an induced representation, according to Eq. (1.7). Each representation is irreducible in $SU(2)$, hence the direct product of the two induced representations is identical to the process of addition of two spin 1/2 angular-momenta.
3. Then, it may be shown that the representation obtained in the direct product space is reducible into irreducible representations in the blocks of dimensions $1 \times 1$ and $3 \times 3$. The first of these may be associated with a scalar field and the second one may be associated with a 3-vector. Both representations correspond to tensors that transform covariantly under the Lorentz group.

We start by defining $U_a^{-1}[L(\Lambda n)], U_a(\Lambda), U_a[L(n)]$ as (non-unitary) matrix realizations belonging to the group $SL(2,\mathbb{C})_a$, acting on states belonging to the Hilbert space $\mathcal{H}_a$, where $a = 1,2$. In this stage, the complete Hilbert space in the spin degrees of freedom is a direct product: $\mathcal{H}_1 \otimes \mathcal{H}_2$.

Step 1 of the procedure described above leads us to define the following transformation matrices:

$$U_{1\otimes 2}^{-1}[L(\Lambda n)]U_{1\otimes 2}(\Lambda)U_{1\otimes 2}[L(n)] =$$
$$(U_1^{-1}[L(\Lambda n)]\otimes U_2^{-1}[L(\Lambda n)])(U_1(\Lambda)\otimes U_2(\Lambda))(U_1[L(n)]\otimes U_2[L(n)]) \quad (2.1)$$

Note that both the $n$ vector and $\Lambda$ are the same in both Hilbert spaces. This fact is crucial as we are dealing with the same physical particle and so we must make the same transformations in all the Hilbert spaces which belong to it. Now, since $\mathcal{H}_1$ and $\mathcal{H}_2$ are two disjoint Hilbert spaces, we can interchange between operators of the different spaces to obtain (Step 2):

$$U_{1\otimes 2}^{-1}[L(\Lambda n)]U_{1\otimes 2}(\Lambda)U_{1\otimes 2}[L(n)]|\sigma_1,\sigma_2\rangle =$$
$$(U_1^{-1}[L(\Lambda n)]U_1(\Lambda)U_1[L(n)]|\sigma_1\rangle)(U_2^{-1}[L(\Lambda n)]U_2(\Lambda)U_2[L(n)]|\sigma_2\rangle) \quad (2.2)$$

We notice in the right hand side of Eq. (2.2) that the first factor is related to $\mathcal{H}_1$ only, while the second factor is related to $\mathcal{H}_2$ only. Therefore, we can treat the two pieces separately using Eq. (1.7) for each of them to obtain two induced representations in the two independent SU(2) manifolds:

$$(U_1^{-1}[L(\Lambda n)]U_1(\Lambda)U_1[L(n)]|\sigma_1\rangle)(U_2^{-1}[L(\Lambda n)]U_2(\Lambda)U_2[L(n)]|\sigma_2\rangle) =$$
$$\left(D_{\sigma_1'\sigma_1}(\Lambda,n)|\sigma_1\rangle\right)\left(D_{\sigma_2'\sigma_2}(\Lambda,n)|\sigma_2\rangle\right) \quad (2.3)$$

In the last expressions (equations 2.2-2.3) we deal with the basis states:

$$|\sigma_1,\sigma_2\rangle \equiv |n;\sigma_1,\sigma_2\rangle = |n,\sigma_1\rangle \otimes |n,\sigma_2\rangle \equiv\equiv |\sigma_1\rangle \otimes |\sigma_2\rangle \quad (2.4)$$

where $\{|\sigma_1\rangle\}$ denote eigenvectors of the angular momentum operator in the $s = \frac{1}{2}$ representation of $SU(2)_a$, projected on the same (arbitrarily chosen) quantization axis, in the $\{\mathcal{H}_a\}$ spaces.

We can shift our attention from the rotation matrices, $\{D^a\}$, to the generators, $\{\frac{1}{2}\vec{\sigma}^a\}$, of the corresponding $\{su(2)_a\}$ algebras. In taking the rotation parameters to be infinitesimal, the direct product can be carried out straightforwardly. This leads us to the total spin operator: $\vec{S}^{(1\oplus 2)} = \frac{1}{2}\vec{\sigma}^{(1)}\otimes 1^{(2)} + 1^{(1)}\otimes\frac{1}{2}\vec{\sigma}^{(2)}$, the eigenvalues of which are used to label the irreducible blocks of the rotation matrices.

Step 3 is the decomposition into irreducible representations, which is executed by performing a similarity transformation: $\vec{S}' = C^{-1}\vec{S}C$. The transformation matrix $C$ takes the original (individual spins) basis into the new (total spin) basis. It is built out of the new basis vectors represented in the old basis, hence C is both orthogonal and has unit determinant, and so: $C^{-1} = C^T = C^\dagger$. The components of the transition matrix are the Clebsch-Gordan coefficients (also called Wigner coefficients):

$$C_{s\sigma;\sigma_1\sigma_2} = C(s_1\sigma_1;s_2\sigma_2|s,\sigma) \equiv C^{ss_1s_2}_{\sigma\sigma_1\sigma_2} \quad (2.5)$$

In this expression: $s_1$ and $s_2$ are the magnitudes of the spins in the fundamental representation, $\sigma_1$ and $\sigma_2$ are their projections on the quantization axis, $s = s_1 \oplus s_2$ is the total spin and $\sigma = \sigma_1 + \sigma_2$ is its projection. Since $s_1 = s_2 = \frac{1}{2}$, we have that $s \in \{0,1\}$.

In going from the old basis to the new one, the representation matrices become block diagonal with $s = 0,1$ irreducible representations in its blocks, in which the states are anti-symmetric (singlet) or symmetric (triplet) under parity transformation, respectively. Therefore, the Hilbert space breaks into subspaces: $\mathcal{H} = \mathcal{H}^0 \oplus \mathcal{H}^1$.

## 2.2 Generalization to Arbitrary Integer and Half-Integer Spins

We now generalize the method of subsection 2.1 in order to obtain higher representations of the little group[2]. This will be done by the following steps:

1. Replace each of the $SL(2, \mathbb{C})$ three factors in the chain operation (Eq. 1.5) by an outer product of N $\{SL(2, \mathbb{C})_a\}$ matrices of the $\frac{1}{2}$ representation, each belonging to a different Hilbert space, $\{\mathcal{H}_a\}$. Each of the three outer products results in a matrix of dimension $2^N$ in the direct product space $\prod_{a=1}^{N} \mathcal{H}_a$ with the underlying symmetry group of $\prod_{a=1}^{N} SL(2, \mathbb{C})_a$.
2. Perform first the simple multiplications between the matrices in each spin-space separately, and only then carry out the outer product of the N resulting matrices. In each spin-space we obtain an induced representation by the method described in subsection 2.1. Each representation $D_a$ is irreducible in $SU(2)_a$, hence the outer product of the induced representations is identical to the process of addition of N spin 1/2 angular-momenta.
3. The representation matrix reduces into a block diagonal form with irreducible representations in the blocks. It may happen that some of the irreducible representations labeled by the index $s \in \{0, \frac{1}{2}, 1, \dots, \frac{N}{2}\}$, will occur several times with the same dimension. Such representations are equivalent to one another and will not be distinguished in our analysis. The highest dimensional irreducible representation is of dimension $(N + 1) \times (N + 1)$. This representation is unique up to a similarity transformation, and so may be associated with a field of spin N/2 in SHP theory. The rest of these irreducible representations are of lower dimensions and are not unique. These can be interpreted (as from our analysis below) as fields with spins $s \in \{0, \frac{1}{2}, 1, \dots, \frac{N}{2}-1\}$.

We now carry out the decomposition in term of the representation matrices. Step 1 is described by a formula which is analogous to Eq. (2.1) of the previous section:

$$U_\Pi^{-1}[L(\Lambda n)]U_\Pi(\Lambda)U_\Pi[L(n)] = (\prod_{a=1}^{N} \otimes U_a^{-1}[L(\Lambda n)])(\prod_{a=1}^{N} \otimes U_a(\Lambda))(\prod_{a=1}^{N} \otimes U_1[L(n)]) \quad (2.6)$$

---

[2] Weinberg[19] has constructed a somewhat similar mapping for tensor fields using the nonunitary decomposition of the Lorentz algebra in order to achieve the Feynman rules for fields of any spin.

Step 2 is described by a formula which is analogous to Eq. (2.2) of the previous section:

$$U_\Pi^{-1}[L(\Lambda n)]U_\Pi(\Lambda)U_\Pi[L(n)]|\{\sigma_a\}\rangle = \prod_{a=1}^{N}\otimes(U_a^{-1}[L(\Lambda n)]U_a(\Lambda)U_a[L(n)]|\sigma_a\rangle) \quad (2.7)$$

and by the following formula, which is analogous to Eq. (2.3) of the previous section:

$$\prod_{a=1}^{N}\otimes(U_a^{-1}[L(\Lambda n)]U_a(\Lambda)U_a[L(n)]|\sigma_a\rangle) = \prod_{a=1}^{N}\otimes(D_{\sigma_{a'}\sigma_a}(\Lambda,n)|\sigma_a\rangle) \quad (2.8)$$

As we did in the lower spin cases, we now identify each multiplicand in the left-hand-side of Eq. (2.8) as a rotation matrix acting in the spin-space and expand it to first order in the angular parameter $\alpha$:

$$D^a = \exp\left(-i\vec{\alpha}\cdot\tfrac{1}{2}\vec{\sigma}^a\right) \approx 1 - i\vec{\alpha}\cdot\tfrac{1}{2}\vec{\sigma}^a \quad (2.9)$$

Step 3 corresponds to the process of addition of N spin 1/2 angular momenta[23]:

$$\prod_{a=1}^{N}\otimes D^a = \prod_{a=1}^{N}\otimes 1^a - i\vec{\alpha}\cdot\sum_{a=1}^{N}\tfrac{1}{2}\vec{\sigma}^a\otimes 1^{N/a} = 1 - i\vec{\alpha}\cdot\vec{S} \quad (2.10)$$

where the notation $N/a$ refers to the indices, out of the $N$ possibilities, which are not equal to a. The decomposition is obtained as a diagonalization of the operators $\vec{S} = \sum_{a=1}^{N}\tfrac{1}{2}\vec{\sigma}^a\otimes 1^{N/a}$ corresponding to the total internal (spin) angular momentum. The three components of this operator-vector are the generators of $SU(2)$, which is the diagonal subgroup of the direct-product group[21]. Block-diagonalization of the D matrices is obtained, as in the previous section, by a change of basis from the direct-product set $|n,\{\sigma_a\}\rangle$ (for $\sigma_a = \pm\tfrac{1}{2}$), to the basis which diagonalizes both the $S_3$ generator and the Casimir operator $S^2 = S_1^2 + S_2^2 + S_3^2$, i.e. to the set of eigenvectors $\{|n,s,\sigma\rangle\}$ (for $-s \leq \sigma \leq +s$).

As a particular but important case, we highlight the states of maximal and minimal total spin. These states do not suffer from the change of basis, so we have: $|S = s, S_3 = \pm s\rangle = |\sigma_1 = \pm\tfrac{1}{2}, \ldots, \sigma_N = \pm\tfrac{1}{2}\rangle$.

The maximal/minimal spin states define the two edges of a scale of $2s+1$ states which all share the same eigenvalue of the Casimir operator and are separated from one another by differences of $\Delta S_3 = 1$. These all represent spin s fields.

# Example - The Spin 3/2 Representation

The formalism described, that generates the spin 3/2 representation, is applicable, for example, for the Rarita-Schwinger field[22]. This is constructed from the composition of three fundamental spins[23] in the induced representation (see Appendix for technical details of the recoupling procedure). In this case we have $a_3 = \tfrac{1}{3}\binom{4}{2} = 2$ options for commutative and associative couplings, $c_3 = 3!\, a_3 = 12$ options for non-commutative but associative couplings and thus there are $d_3 = \tfrac{c_3}{2^2} = 3!! = 3$ options for non-commutative and non-associative schemes, given by:

$$\begin{cases} |(\vec{A}\vec{B}\vec{C})[(ab)_d c]_{s\sigma}\rangle = \sum_{\alpha\beta} C^{abd}_{\alpha\beta,\alpha+\beta} C^{dcs}_{\alpha+\beta,\sigma-\alpha-\beta,\sigma} |a,\alpha;b,\beta;c,\sigma-\alpha-\beta\rangle \\ |(\vec{B}\vec{C}\vec{A})[(bc)_e a]_{s\sigma}\rangle = \sum_{\beta\gamma} C^{bce}_{\beta\gamma,\beta+\gamma} C^{eas}_{\beta+\gamma,\sigma-\beta-\gamma,\sigma} |a,\sigma-\beta-\gamma;b,\beta;c,\gamma\rangle \quad (2.11) \\ |(\vec{C}\vec{A}\vec{B})[(ca)_f b]_{s\sigma}\rangle = \sum_{\gamma\alpha} C^{caf}_{\gamma\alpha,\gamma+\alpha} C^{fbs}_{\gamma+\alpha,\sigma-\gamma-\alpha,\sigma} |a\alpha;b\sigma-\gamma-\alpha;c,\gamma\rangle \end{cases}$$

In these expressions and the following $a, b, c$ refer to the eigenvalues of the angular momenta $\vec{A}, \vec{B}, \vec{C}$ of the individual spins to be added, whereas $d, e, f$ refer to the eigenvalues of the angular momenta in the intermediate states of the composition.

The transition from one nonequivalent scheme to the other is given by $\binom{3}{2} = 3$ Racah coefficients (see Appendix) that are given by:

$$\begin{cases} \langle (\vec{A}\vec{B}\vec{C})[(ab)_d c]_s | (\vec{B}\vec{C}\vec{A})[(bc)_e a]_s \rangle = (-1)^{a+e-s} W(absc;de) \\ \langle (\vec{B}\vec{C}\vec{A})[(bc)_e a]_s | (\vec{C}\vec{A}\vec{B})[(ca)_f b]_s \rangle = (-1)^{b+f-s} W(bcsa;ef) \quad (2.12) \\ \langle (\vec{C}\vec{A}\vec{B})[(ca)_f b]_s | (\vec{A}\vec{B}\vec{C})[(ab)_d c]_s \rangle = (-1)^{c+d-s} W(casb;fd) \end{cases}$$

The rest of the $c_3 = 12$ transition coefficients can be obtained from those shown in Eq. (2.11) by transpositions, which give rise to phases of the form:

$$|\cdots[(x)_a(y)_b]_d \cdots\rangle = (-1)^{a+b-d} |\cdots[(y)_b(x)_a]_d \cdots\rangle \quad (2.13)$$

An alternative way of adding the 3 spin 1/2 fundamental representations is by the use of the so-called "democratic coupling"[23]. In this method, it is noticed that the triple-product operator $K \equiv (\vec{A} \times \vec{B}) \cdot \vec{C}$ commutes with all the generators of the individual $SU(2)$ groups. It is thus possible to simultaneously diagonalize the six Hermitian operators:

$$\{A^2, B^2, C^2, K, S^2, S_3\} \quad (2.14)$$

The tensor product space $\mathcal{H}_A \otimes \mathcal{H}_B \otimes \mathcal{H}_C = \mathcal{H}(a,b,c)$ is mapped into itself by the action of $K$ and $\vec{S}$, and it must be possible to split $\mathcal{H}(a,b,c)$ into $2s+1$ dimensional irreducible subspaces with respect to $\vec{S}$, where these subspaces are spanned by basis vectors of the form:

$$|k(abc)_{s\sigma}\rangle = \sum_{\sum_i \sigma_i = \sigma} C^{abcs}_{\sigma_1 \sigma_2 \sigma_3 \sigma}(k) |a,\sigma_1\rangle |b,\sigma_2\rangle |c,\sigma_3\rangle \quad (2.15)$$

In this expression, $k$ is the eigenvalue of $K$, and it serves to distinguish fully between those subspaces of $\mathcal{H}(a,b,c)$ which are carrier spaces of the same representation $D^s$ of the $SU(2)$ diagonal subgroup of $SU(2) \otimes SU(2) \otimes SU(2)$. The coefficients occurring in Eq. (2.15) are Wigner coefficients for the complete reduction of $SU(2) \otimes SU(2) \otimes SU(2)$ in its diagonal subgroup.

In this method, the three spins are treated alike in the coupling. In our case, we have equal values of the spins $a = b = c = \frac{1}{2}$, and thus the eigenvectors $|k\left(\frac{1}{2},\frac{1}{2},\frac{1}{2},\right)_{s\sigma}\rangle$ belong to subspaces that are irreducible with respect to the action of the symmetric group $\mathcal{S}_3$. This has the consequence of assigning definite statistics under the exchange of the spins[14].

# 3 Construction of Tensors and Spinors in the SHP

## 3.1 Fundamental Spinor Fields

In the SHP framework, one can construct 2-dimensional spinors by projecting Eq. (1.5) on a general ket $|\varphi\rangle$ and complex-conjugating it, leading to wave-functions, $\varphi_{n,\sigma}$, which transform according to:

$$\langle n, \sigma | \varphi \rangle = D_{\sigma'\sigma}(\Lambda, n) \langle \Lambda^{-1} n, \sigma' | \varphi \rangle \quad (3.1)$$

or

$$\varphi_{n,\sigma}(x) = D_{\sigma'\sigma}(\Lambda, n) \varphi_{\Lambda^{-1}n,\sigma}(\Lambda^{-1} x) \quad (3.1')$$

As remarked above, there are two fundamental inequivalent representations of $SL(2, \mathbb{C})$; we have been working with one of them up to now, but we shall point out here a basis for further generalization. The two kinds of representations are distinguished by the use of the helicity operator, $\sigma^\mu n_\mu$. The first/second type of fundamental representations of the covering group of $\ell_+^\uparrow$ is spanned by the 4 linearly independent quaternionic matrices[12]:

$$\begin{cases} \sigma_\mu \leftrightarrow (+\mathbf{1}, \vec{\sigma}) \; basis \; of \; SL(2, \mathbb{C}) \\ \overline{\sigma_\mu} \leftrightarrow (-\mathbf{1}, \vec{\sigma}) \; basis \; of \; \overline{SL(2, \mathbb{C})} \end{cases} \quad (3.2)$$

where $\mathbf{1}$ is the $2 \times 2$ unit matrix while the spacial components are the three Pauli matrices.

Horwitz and Arshansky[13] constructed Dirac type wave-functions as follows. For any Lorentz boost $L(n) \in SL(2, \mathbb{C})$ acting on spinor wave-functions as $U[L(n)]\varphi_{n,\sigma}(x)$, there exists a Lorentz boost $\bar{L}(n) \in \overline{SL(2, \mathbb{C})}$ of the second kind of representations acting on spinor wave-functions of the form: $U[\bar{L}(n)]\chi_{n,\sigma}(x)$. Therefore, we can describe a spin 1/2 particle using a 4-component spinor with components depending on the two 2-component fundamental spinors, $\varphi_{n,\sigma}(x)$ and $\chi_{n,\sigma}(x)$, in the following way[13]:

$$\psi_n(x) = \tfrac{1}{\sqrt{2}} \begin{pmatrix} +1 & +1 \\ -1 & +1 \end{pmatrix} \begin{pmatrix} U[L(n)]\varphi_n(x) \\ U[\bar{L}(n)]\chi_n(x) \end{pmatrix} \quad (3.3)$$

The normalization of such a spinor would depend on the fundamental spinors' norms in the following way:

$$\langle \psi_n | \psi_n \rangle = \mp \int \bar{\psi}_n^*(x) \gamma \cdot n \psi_n(x) d^4 x = \int (|\varphi_n(x)|^2 + |\chi_n(x)|^2) \, d^4 x \quad (3.4)$$

where $\{\gamma^\mu\}$ are the Dirac matrices, the $\mp$ corresponds to $n^\mu$ in the positive/negative light-cone, respectively. This equation generalizes the normalization condition for the case of spin 1/2.

The spinor wave-function $\psi_n$ is an element of the Hilbert space $\mathbb{L}^2(\mathbb{R}^{3,1}, d^4x; \mathbb{R}^3, d^3n) \times \mathbb{C}^4$. It transforms, under a Lorentz transformation $\Lambda \in SO(3,1)$ in the configuration space and in the $n$-space, according to:

$$\psi_n(x) = S(\Lambda)\psi_{\Lambda^{-1}n}(\Lambda^{-1} x) \quad (3.5)$$

where the (non-unitary) operator $S(\Lambda) = e^{-i\omega_{\mu\nu}\Sigma^{\mu\nu}}$ depends on a 6-parameter (3 rotations + 3 boosts) skew-symmetric matrix $\omega_{\mu\nu}$, while $\Sigma^{\mu\nu} \equiv [\gamma^\mu, \gamma^\nu]$ is the realization of the Lorentz group generators in the $\frac{1}{2}$ representation.

The Dirac operator, $\gamma \cdot p$, is not Hermitian in the invariant scalar product associated with Eq. (3.4), but it can be decomposed into Hermitian and anti-Hermitian parts:

$$\begin{cases} K_L = \frac{1}{2}\big(\gamma \cdot p + (\gamma \cdot n)(\gamma \cdot p)(\gamma \cdot n)\big) = -(p \cdot n)(\gamma \cdot n) \\ K_T = \frac{1}{2}\gamma^5\big(\gamma \cdot p + (\gamma \cdot n)(\gamma \cdot p)(\gamma \cdot n)\big) = -2i\gamma^5(p \cdot K)(\gamma \cdot n) \end{cases} \quad (3.6)$$

where $K^\mu \equiv \Sigma^{\mu\nu} n_\nu$ and $\gamma^5 = i\gamma^0\gamma^1\gamma^2\gamma^3$ is the chiral Dirac matrix, which anti-commutes with the vector Dirac matrices and obeys: $(\gamma^5)^2$. It is easy to check that:

$$K_L^2 = (p \cdot n)^2, K_T^2 = p^2 + (p \cdot n)^2 \quad (3.7)$$

Thus, a spin 1/2 particle may be described in the SHP theory by the free Stueckelberg Hamiltonian:

$$K_{\frac{1}{2}}(x,p) = \frac{1}{2M}(K_T^2 - K_L^2) = \frac{p^2}{2M} \quad (3.8)$$

The n-vector gives rise to a projected set of Dirac matrices:

$$\gamma_n^\mu \equiv \gamma_\lambda h^{\lambda\mu} \quad (3.9)$$

where:

$$h^{\lambda\mu} \equiv g^{\lambda\mu} + n^\lambda n^\mu \quad (3.10)$$

In terms of the new Dirac matrices, the infinitesimal generators of Lorentz transformations are given by[13]:

$$\Sigma_n^{\mu\nu} \equiv \frac{i}{4}[\gamma_n^\mu, \gamma_n^\nu] = \Sigma^{\mu\nu} + K^\mu n^\nu - n^\mu K^\nu \quad (3.11)$$

These generators are Hermitian with respect to the inner-product defined in Eq. (3.4). Moreover, they give rise, together with the $K^\mu$ operators defined above, to an algebra in the "projected space" endowed with the projected metric $h$. One can check the following commutation relations, which define this algebra[13]:

$$\begin{cases} [K^\mu, K^\nu] = -i\Sigma_n^{\mu\nu} \\ [\Sigma_n^{\mu\nu}, K^\lambda] = -i\big(h^{\nu\lambda}K^\mu - h^{\mu\lambda}K^\nu\big) \\ [\Sigma_n^{\mu\nu}, \Sigma_n^{\lambda\rho}] = -i\big(h^{\nu\lambda}\Sigma_n^{\mu\rho} + h^{\rho\mu}\Sigma_n^{\nu\lambda} - h^{\mu\lambda}\Sigma_n^{\nu\rho} - h^{\rho\nu}\Sigma_n^{\lambda\mu}\big) \end{cases} \quad (3.12)$$

Since $K^\mu n_\mu = \Sigma^{\mu\nu} n_\mu n_\nu = 0$, there are only 3 independent $K^\mu$, hence only 3 independent $\Sigma_n^{\mu\nu}$. These matrices are the covariant form of the Pauli matrices. Equation (3.12) constitutes a full representation of the Lorentz group.

Returning to Eq. (3.8), although this expression for the spin 1/2 Hamiltonian is numerically identical to the free Hamiltonian of a spinless particle (see section 1), the structure necessary for the construction of the Hamiltonian for a spin 1/2 particle is contained in the operators $K_L$ and $K_T$. In the presence of an electromagnetic field, their gauge invariant forms satisfy the commutation relation:

$$i[K_T, K_L] = -ie\gamma^5(K^\mu n^\nu - n^\mu K^\nu)f_{\mu\nu} \quad (3.13)$$

Where $f_{\mu\nu} \equiv \partial_\mu a_\nu - \partial_\nu a_\mu$ is the electromagnetic field strength tensor, and $a^\mu(x)$ is the electromagnetic 4-potential, which is introduced to the theory by the minimal substitution: $p^\mu - ea^\mu$. The spin 1/2 Hamiltonian becomes then becomes:

$$K_{\frac{1}{2}}(x.p) = \frac{1}{2M}(K_T^2 - K_L^2) = \frac{(p - \frac{e}{c}a)^2}{2M} + \frac{e}{2Mc}\Sigma_n^{\mu\nu} f_{\mu\nu} \quad (3.14)$$

The obtained Hamiltonian is a second order analog of the Dirac Hamiltonian. In the special frame defined by $n = n_0$, one obtains: $\Sigma_{n_0}^{ij} = \epsilon^{ijk}\sigma_k$ (i, j and k are spacial indices). In this frame the interaction is that of ordinary spin coupling to magnetism in the usual way with fine structure coupling $\alpha = \frac{e^2}{\hbar c}$, corresponding to the Dirac gyromagnetic moment with value 2. Since the electric coupling is zero, the problem of the presence of the non-Hermitian term in the second order Dirac equation for the electric coupling is removed[3].

The 1-loop correction to the magnetic moment of a particle evolving with the Hamiltonian (3.14) was calculated by Bennett [24], resulting in a value compatible with the one obtained by renormalization of the vertex function in QFT[25].

In relativistic scattering theory, the asymptotic states of a spin 1/2 particle can be decomposed according to the projection operators:

$$\begin{cases} P_\pm = \frac{1}{2}(\mathbf{1} \mp \gamma \cdot n) \\ P_{E\pm} = \frac{1}{2}\left(\mathbf{1} \mp \frac{p \cdot n}{|p \cdot n|}\right) \\ P_{n\pm} = \frac{1}{2}\left(\mathbf{1} \mp \frac{2i\gamma^5 p \cdot K}{\sqrt{p^2 + (p \cdot n)^2}}\right) \end{cases} \quad (3.15)$$

where again the $\pm$ sign is according to the direction of $n^\mu$ with respect to the light-cone. In the frame with $n = n_0$, one obtains the simple forms:

$$\begin{cases} P_\pm \to \frac{1}{2}(\mathbf{1} \mp \gamma^0) \\ P_{E\pm} \to \frac{1}{2}(\mathbf{1} \mp sgn(E)) \\ P_{n\pm} \to \frac{1}{2}\left(\mathbf{1} \mp \frac{\vec{\sigma} \cdot \vec{p}}{|\vec{p}|}\right) \end{cases} \quad (3.16)$$

Therefore, one retains the definitions of projection operators of the parity, energy-branch and helicity operators as formulated in the Dirac theory.

Finally, note that the discrete symmetries of the SHP wave-function are:

---

[3] In his calculation of the anomalous moment of the electron, Schwinger[15], using a formalism closely related to that of Stueckelberg[1], set the electric field equal to zero.

$$\begin{cases} \psi^C(t,\vec{x};n_0,\vec{n};\tau) = i\gamma^2\psi^*(t,\vec{x};n_0,\vec{n};-\tau) \\ \psi^P(t,\vec{x};n_0,\vec{n};\tau) = \gamma^0\psi^*(t,-\vec{x};n_0,-\vec{n};\tau) \\ \psi^T(t,\vec{x};n_0,\vec{n};\tau) = i\gamma^1\gamma^3\psi^*(-t,\vec{x};-n_0,\vec{n};-\tau) \quad (3.17)\\ \psi^{CP}(t,\vec{x};n_0,\vec{n};\tau) = i\gamma^2\gamma^0\psi^*(t,-\vec{x};n_0,-\vec{n};-\tau) \\ \psi^{CPT}(t,\vec{x};n_0,\vec{n};\tau) = i\gamma^5\psi^*(-t,-\vec{x};-n_0,-\vec{n};\tau) \end{cases}$$

According to this, the CPT-conjugate wave-function moves backwards in space-time relative to the motion of the original wave-function, but moves forward in $\tau$. For a wave-packet with $E < 0$ components, which moves backwards in t as $\tau$ goes forward, it is the CPT wave-function which moves forward with an opposite charge, i.e. the observed antiparticle. No Dirac sea is required for the consistency of the theory, since unbounded transitions to the $E < 0$ domain are prevented by the conservation of K.

## 3.2 Vector Fields

The method outlined in section 2 suggests that we may be able to build a vector field of spin 1. These fields may be obtained from the multiplication of spinors of the fundamental representation of the little group, but the general spin 1 tensor has a more general form, which only transforms irreducibly in the same way as the direct product of spinors does.

The components of a 4-vector $a = (a^0, a^1, a^2, a^3)$ can be viewed as components of a quaternionic number, thus allowing the definition of an associated $SL(2,\mathbb{C})$ matrix[12]:

$$\mathcal{A} \equiv \tfrac{1}{\sqrt{2}}\sigma_\mu a^\mu = \tfrac{1}{\sqrt{2}}\begin{pmatrix} a^0 + a^3 & a^1 + ia^2 \\ a^1 - ia^2 & a^0 - a^3 \end{pmatrix} \quad (3.18)$$

Note that the determinant of the $\mathcal{A}$-matrix is proportional to the norm of the $a$-vector:

$$det(\mathcal{A}) = \tfrac{1}{2}[(a^0)^2 + (a^1)^2 + (a^2)^2 + (a^3)^2] = \tfrac{1}{2}a_\mu a^\mu \quad (3.19)$$

The invariance of the determinant under Lorentz transformations in the Hilbert space leads to the invariance of the norm under Lorentz transformations in configuration space.

One can relate the $SL(2,C)$ matrix we have obtained to the Lorentz four vectors as follows. If one attaches a vector label, n, to the $\mathcal{A}$-matrix of Eq. (3.18), one can investigate the response of this vector to an application of a Lorentz transformation. First note that an application of a Lorentz boost affects the n-vector. By extrapolation, there exists an origin, $n_0$, such that any $SL(2,\mathbb{C})$ matrix, $\mathcal{A}_n$, can be obtained from $\mathcal{A}_{n_0}$ by applying a suitable Lorentz boost:

$$\mathcal{A}_n = U[L(n)]\mathcal{A}_{n_0}U^{-1}[L(n)] \quad (3.20)$$

Therefore, the $n$-vector is the stability vector of the induced little group.

Application of a Lorentz transformation on such form leads to:

$$U(\Lambda)\mathcal{A}_n U^{-1}(\Lambda) = U(\Lambda)U[L(n)]\mathcal{A}_{n_0}U^{-1}[L(n)]U^{-1}(\Lambda) = \mathcal{A}_{\Lambda n}\left(D^{\frac{1}{2}}(\Lambda,n)\otimes D^{\frac{1}{2}}(\Lambda,n)\right)$$
(3.21)

where in the first equality we used Eq. (3.20), in the second we used the definition of the induced representation in each side of the $\mathcal{A}$-matrix and boosted the $n$-vector to its transformed value on the orbit. Using the reduction of direct product of representations given in subsection 2.2, we are able to decompose the 4-vector into its rotational-scalar and rotational-vector parts:

$$U(\Lambda)\mathcal{A}_n U^{-1}(\Lambda) = \mathcal{A}_{\Lambda n}(D^0(\Lambda,n)\oplus D^1(\Lambda,n)) \quad (3.22)$$

Thus, there is an equivalence between the on-orbit tensor $\mathcal{A}_n$ which decomposes into irreducible elements of $SO(3,1) \simeq SU(2)$, as opposed to the off-orbit $\mathcal{A}_{n_0}$ which transforms irreducibly as an element of $SL(2,\mathbb{C})$.

It is easy to see that applying the transformation $\mathcal{A}_n \to U^{-1}[L(n)]\mathcal{A}_n U[L(n)]$ results in a tensor with the usual Lorentz properties for the 4 dimensional representation, mixing the space and time indices of $\mathcal{A}$ just as in the transition from the $SL(2,\mathbb{C})$ representations of a spinor to the Dirac form (leaving a transformation law independent of $n$).

The vector field that we have just constructed may satisfy certain dynamical equations, depending on the physical content that is inherent in the field. In many contexts in contemporary modern physics, vector fields emerge as connection forms that arise as a result of $U(1)$ gauge transformations of the following form (here, we concentrate on the Abelian case):

$$a^\mu \to a^\mu + \partial^\mu \lambda \quad (3.23)$$

where $\lambda(x)$ is some arbitrary differentiable function. These symmetries constrain the possible forms of the Lagrangian of the theory, which yields second order field equations, into the following set: $\mathcal{L} = \{f_{\mu\nu}f^{\mu\nu}, f_{\mu\nu}\tilde{f}^{\mu\nu}\}$, with the usual definitions for the field strength tensor $f_{\mu\nu} \equiv \partial_\mu a_\nu - \partial_\nu a_\mu$ and its dual tensor $\tilde{f}^{\mu\nu} \equiv \frac{1}{2}\varepsilon^{\mu\nu\rho\sigma}f_{\mu\nu}$. Taking the gauge invariance of $\mathcal{L}$ into account allows one to make concrete calculations using a specific gauge that fits problem at hand. The most common gauges in particle-physics are the Lorentz gauge and the Coulomb gauge, both of which can be written in the form: $\partial_\mu a^\mu = g$, with $g(x)$ being some function of space-time coordinates.

We shall now cast the gauge symmetry and the dynamical equations that the vector field may satisfy, into a form that can be related to the spin fields in the induced representation. For that purpose we shall start with writing the 4-derivative in a matrix form:

$$\mathcal{D} \equiv \tfrac{1}{\sqrt{2}}\bar{\sigma}_\mu \partial^\mu = \tfrac{1}{\sqrt{2}}\begin{pmatrix} \partial^0 + \partial^3 & \partial^1 + i\partial^2 \\ \partial^1 - i\partial^2 & \partial^0 - \partial^3 \end{pmatrix} (3.24)$$

Applying this derivative-matrix on the $\mathcal{A}$-matrix which contains the vector field components, we obtain a $2 \times 2$ matrix which contains the gauge condition and the field strength components:

$$\mathcal{D}\cdot\mathcal{A} = \tfrac{1}{2}\bar{\sigma}_\mu \sigma_\nu \partial^\mu a^\nu = \tfrac{1}{2}\begin{pmatrix} \partial\cdot a + f^{03} + if^{12} & f^{01} - if^{02} + if^{23} + f^{31} \\ f^{01} + if^{02} + if^{23} - f^{31} & \partial\cdot a - f^{03} - if^{12} \end{pmatrix} \quad (3.25)$$

This expression can be further manipulated into a form which exhibits the electric and magnetic fields explicitly:

$$\mathcal{D}\cdot\mathcal{A} = \tfrac{1}{2}(\partial\cdot a)\mathbf{1} + \tfrac{1}{2}(\vec{\varepsilon} + i\vec{b})\cdot\vec{\sigma} = \tfrac{1}{2}\begin{pmatrix} \partial\cdot a + \varepsilon^3 + ib^3 & \varepsilon^1 - i\varepsilon^2 + ib^1 + b^2 \\ \varepsilon^1 + i\varepsilon^2 + ib^1 - b^2 & \partial\cdot a - \varepsilon^3 - ib^3 \end{pmatrix} \quad (3.25')$$

Note that the trace of this matrix yields the gauge-fixing relation:

$$tr(\mathcal{D}\cdot\mathcal{A}) = \partial\cdot a \quad (3.26)$$

Furthermore, the determinant of the same matrix yields an expression that combines the two possible forms of the gauge invariant free Lagrangians:

$$det(\mathcal{D}\cdot\mathcal{A}) = \tfrac{1}{4}[\partial\cdot a + f_{\mu\nu}f^{\mu\nu} + f_{\mu\nu}\tilde{f}^{\mu\nu}] \quad (3.27)$$

This form constitutes a realization of the $0\oplus 1$ representation of the Lorentz group[26]. The fact that Eq. (3.27) contains both $\{f_{\mu\nu}f^{\mu\nu}, f_{\mu\nu}\tilde{f}^{\mu\nu}\}$ suggests that our analysis may be applicable not only to Maxwell-like field theories like QED, but also to non-Abelian theories like QCD where $f_{\mu\nu}\tilde{f}^{\mu\nu}$ does not vanish. This will be studied in a future work.

In the quaternionic formalism that was just presented, Hamilton's principle may be cast in the following form:

$$\delta S = 0 \text{ with } S = \int d^4x\{det(\mathcal{D}\cdot\mathcal{A}) + \beta[tr(\mathcal{D}\cdot\mathcal{A}) - \partial\cdot a]\} \quad (3.28)$$

where $\beta$ is a Lagrange multiplier. Variation of this action leads to dynamical equations that are satisfied by the vector field, under the constraint imposed by the gauge:

$$\begin{cases} \partial_\mu f^{\mu\nu} = 0 \\ \partial_\mu a^\mu = g \end{cases} \quad (3.29)$$

Before leaving this topic, we shall note that the construction the vector field could have been done using both fundamental representations. There are in total four options of combinations, according to which the vector field in the SHP theory may transform like:

$$\begin{cases} \varphi_n^{(1)\dagger}\otimes\varphi_n^{(2)} \\ \varphi_n^{(1)\dagger}\otimes\chi_n^{(2)} \\ \chi_n^{(1)\dagger}\otimes\varphi_n^{(2)} \\ \chi_n^{(1)\dagger}\otimes\chi_n^{(2)} \end{cases} \quad (3.30)$$

## 3.3 Higher Rank Tensor and Spinor Fields

In this section, we are using the mapping from a Lorentz 4-vector into its associated $SL(2,\mathbb{C})$ matrix for the construction of tensors and spinors of rank which is higher than 1. This is done iteratively, using an increasing number of spin 1/2 irreducible representations as the building blocks of the construction. An even number of $SL(2,\mathbb{C})$ blocks induces tensor indices, while an odd number of them induces spinor indices.

Tensors of rank higher than 1 may be obtained by considering mathematical objects of many indices each of which transforms as a Lorentz four vector. For example, if one wishes to construct a rank-2 tensor, one may consider the 2-indexed tensor, $t^{\mu\nu}$. Since $\mu$ and $\nu$ are Lorentz indices, $t^{\mu\nu}$ transforms under the Lorentz group as $a^\mu b^\nu$, with $a^\mu, b^\nu \in \mathbb{R}^{3,1}$. The direct product of the 4-vectors can be cast into a matrix form in the covering group of $\ell_+^\uparrow$. We build $\mathcal{A}, \mathcal{B}$, the corresponding $SL(2,\mathbb{C})$ matrices of the 4-vectors according to Eq. (3.18) and take the direct product of these matrices. Thus the identification $a^\mu b^\nu \to \mathcal{A} \otimes \mathcal{B}$ means that there exists an $SL(2,\mathbb{C}) \otimes SL(2,\mathbb{C})$ matrix, $\mathcal{T}$, that is related to the tensor $t^{\mu\nu}$ and transforms under the Lorentz group as $\mathcal{A} \otimes \mathcal{B}$.

Using the relation between the 4-vector components and the associated $SL(2,\mathbb{C})$ matrix components:

$$\begin{cases} a^0 = \frac{1}{\sqrt{2}}(A^{11} + A^{22}) \\ a^1 = \frac{1}{\sqrt{2}}(A^{11} - A^{22}) \\ a^2 = \frac{1}{\sqrt{2}i}(A^{12} + A^{21}) \\ a^3 = \frac{1}{\sqrt{2}}(A^{12} - A^{21}) \end{cases} \quad (3.31)$$

one can express the rank-2 tensor components in terms of the associated $SL(2,\mathbb{C}) \otimes SL(2,\mathbb{C})$ matrix components. For example:

$$t^{\mu=0,\nu=0} = \frac{1}{\sqrt{2}}(T^{\mu=0,11} + T^{\mu=0,22}) = \frac{1}{2}(T^{11,11} + T^{22,11} + T^{11,22} + T^{22,22}) \quad (3.32)$$

and in a similar way:

$$t^{\mu=0,\nu=1} = \frac{1}{\sqrt{2}}(T^{\mu=0,11} - T^{\mu=0,22}) = \frac{1}{2}(T^{11,11} + T^{22,11} - T^{11,22} - T^{22,22}) \quad (3.33)$$

In general, a mixed tensor $T^{\mu_1\ldots\mu_m}_{\nu_1\ldots\nu_n}$ with $m + n = s$ is of rank $s$ if it transforms like $\prod_{k=1}^{m}\prod_{l=1}^{n} a^{\mu_k} b_{\nu_l}$, with $\{a^{\mu_k}\}\{b^{\nu_l}\} \in \mathbb{R}^{3,1}$. Such a tensor may be obtained by the direct product of s matrices of $SL(2,\mathbb{C})$ on the orbit, thus rendering the tensor spin s in a covariant way.

Spinors of rank higher than a Dirac spinor may be obtained by considering mathematical objects of many indices each of which represents a Lorentz fundamental spinor, i.e. transforming according to Eq. (3.1). For example, if one wishes to construct an $s = \frac{3}{2}$ spinor, one can consider the object $\psi_\alpha^\mu$, with $\mu$ as a Lorentz index and with $\alpha$ as a Dirac index[27]. This object transforms under the Lorentz group as $\psi_\alpha a^\mu$, with $\psi_\alpha$ a Dirac spinor and with $a^\mu \in \mathbb{R}^{3,1}$ a Lorentz vector. Note that such a Lorentz vector may by itself be constructed as described in the previous section, thus making evident the fact that an s=3/2 spinor can be constructed out of three fundamental representations of the induced little group.

The same considerations apply to spinors of higher rank. In general, a mixed spinor $\psi^{\mu_1...\mu_m}_{\alpha_1...\alpha_n}$ with $m + n = s$ is of rank $s$ if it transforms like $\prod_{k=1}^{m} \prod_{l=1}^{n} a^{\mu_k} \psi_{\alpha_l}$, with $a^{\mu_k} \in \mathbb{R}^{3,1}$ and $\psi_{\alpha_l}$ as fundamental spinor variables.

A spinor-field is obtained if one replaces the vectors $\{a^{\mu_k}\}$ with vector-fields and the fundamental spinors $\{\psi_{\alpha_l}\}$ with spinor fields, each belonging to appropriate Hilbert spaces.

# 4 Pauli-Lubanski Vector in the SHP Theory

In this section we discuss a covariant Pauli-Lubanski vector, $W^\mu$, which in the rest frame of the particle carries the physical internal angular momentum of the particle and for which the invariant $g_{\mu\nu} W^\mu W^\nu$ serves as the second Casimir operator for the Poincaré group[28].

The angular momentum operator imbedded in this definition generates rotations in the hyperplane orthogonal to the stability vector $n$ labeling the point on the orbit of the induced representation.

In order to construct the relations (1.6), the unitary operator representing the Lorentz group must be of the following form:

$$M_n^{\mu\nu} = x^\mu p^\nu - x^\nu p^\mu - i\left(n^\mu \frac{\partial}{\partial n_\nu} - n^\nu \frac{\partial}{\partial n_\mu}\right) \quad (4.1)$$

This operator is Hermitian with respect to the full scalar product, including integration over the measure $d^4 n \times \delta(n^2 + 1)$, i.e. on $\frac{d^3 n}{2n^0}$. However, we are interested here in the dynamical spin of the particle at a point $n^\mu$ on the orbit of the induced representation. We therefore define, as for the theory of spin 1/2 particles (see Subsection 3.1):

$$M_n^{\mu\nu} = M^{\mu\nu} + K^\mu n^\nu - K^\nu n^\mu \quad (4.2)$$

where

$$K^\mu = M^{\mu\nu} n_\nu \quad (4.3)$$

The operator $M_n^{\mu\nu}$ acts as an $SU(2)$ rotation in a spacelike plane perpendicular to the timelike vector $n$, which is identified with the physical angular momentum of the particle[14]. To prove this statement we first compute:

$$[M_n^{\mu\nu}, M_n^{\kappa\lambda}] = i\left(h^{\mu\kappa} M_n^{\nu\lambda} - h^{\mu\lambda} M_n^{\nu\kappa} - h^{\nu\kappa} M_n^{\mu\lambda} + h^{\nu\lambda} M_n^{\mu\kappa}\right) \quad (4.4)$$

where the $h$ tensor is the metric of equation (3.10), which projects all vectors to a subspace orthogonal to $n^4$. In the special case for which $n^\mu = (+1,0,0,0)$, Eq. (4.4) reduces to the commutation relations of ordinary three dimensional angular momentum. Therefore, the operator $M_n^{\mu\nu}$ is a covariant form of the Lie generators of $SU(2)$, valid for any value of spin.

---

[4] Note that the projection $h^{\mu\nu}$ effectively brings the metric into a three dimensional Euclidean space with signature (+++) by the operation $h_{\mu\nu} g^{\nu\lambda} h_{\lambda\kappa} = h_{\mu\kappa}$.

Furthermore, since $n_\mu K^\mu = n_\mu M^{\mu\nu} n_\nu = 0$, there are just three independent $K$ components and three independent $M_n$ components. We therefore see that the operators $M_n^{\mu\nu}$ generate rotations in a hyperplane orthogonal to the timelike vector $n$. These are identified as the physical angular momentum components of the particle.

Since $M_n^{\mu\nu} n_\nu$ is identically zero, it is clear that $M_n^{\mu\nu}$ rotates the position vector $x^\lambda$ in a plane perpendicular to $n_\nu$. When the rotation is taken to be infinitesimal, the following commutation relation arises:

$$[M_n^{\mu\nu}, x^\lambda] = ih^{\mu\lambda}x^\nu - ih^{\nu\lambda}x^\mu + i(h^{\mu\lambda}n^\nu + h^{\nu\lambda}n^\mu)(x \cdot n) \quad (4.5)$$

We now define the Pauli-Lubanski operator[26] in our context:

$$(W_\mu)_n = \tfrac{1}{2}\epsilon_{\mu\nu\kappa\lambda} M_n^{\nu\kappa} p^\lambda \quad (4.6)$$

We may easily demonstrate that this operator is Hermitian[5]. Since the commutator of $M_n^{\nu\kappa}$ with $p^\lambda$ has the same form as with $x^\lambda$ (with $x^\lambda$ replaced with $p^\lambda$), and since $\epsilon_{\mu\nu\kappa\lambda}$ is totally anti-symmetric, we have: $\epsilon_{\mu\nu\kappa\lambda}[M_n^{\nu\kappa}, p^\lambda] = 0$. We can therefore define the Casimir operator on the orbit:

$$C_n \equiv (W_\mu)_n (W^\mu)_n \quad (4.7)$$

This operator commutes with the first Poincaré Casimir $p_\mu p^\mu$, corresponding to the mass of the particle (not necessarily a constant of the motion in the relativistic dynamics of SHP). If the momentum of the particle takes on the a value parallel to $n^\mu$, the Pauli-Lubanski operator that we have defined then coincides with the covariant relativistic generalization of the intrinsic physical angular momentum on the orbit. In this case, a Lorentz transformation to the rest frame for which $n^\mu = (+1,0,0,0)$ brings $M_n^{\mu\nu}$ explicitly to the form of a generator of $SU(2)$, as remarked in connection to Eq. (4.4). Note furthermore, that we note that in an asymptotic state with well defined wave-packet, if $n^\mu$ and $p^\mu$, in the sense that $n^\mu \cong \frac{p^\mu}{m}$ (for $m = \sqrt{-p_\mu p^\mu}$), the derivative of the little group stability vector due to Lorentz transformations would be proportional to:

$$\frac{\partial n^\mu}{\partial p_\nu} = \frac{1}{m}\left(g^{\mu\nu} + \frac{p^\mu p^\nu}{m^2}\right) \quad (4.8)$$

which projects to a vector orthogonal to $x^\mu \sim p^\mu$. The state of this wave-packet, on which we can expect its modulation by the action of the little group and its derivative to have only a small effect on the conclusion, then forms, in the construction of $\langle x^\mu \rangle$, an (approximate) expectation value of the operator in (4.8). In such an asymptotic state, for which the momentum is fairly sharp, the expected value of this operator would be very small. In this way, the approximate alignment of $n^\mu$ and $p^\mu$ would retain the required covariance of the expectation value of $x^\mu$.

Thus, the stability vector could be thought of as defining a frame (for example, for the Stern-Gerlach measurement of the spin of an asymptotic state) in which the intrinsic spin,

---

[5] The second term of (4.5) is Hermitian on integration over the $n^\mu$ foliation, an intrinsic part of the scalar product on the full Hilbert space

corresponding to the physical angular momentum of the particle, as it occurs explicitly in the induced representation, can be directly measured[29].

We have therefore constructed a form of the Poincaré group on the orbit of the induced representation, algebraically equivalent to a covariant generalization of the Galilean group attached to a point n of the induced representation, retaining the mass of the particle as an observable.

# 5 Conclusions and Discussion

We have shown that any rank tensor containing integer or half integer can be constructed on the induced representation orbit, using a time-like stability vector. This method has implications for the statistics of the fields constructed in this way as is clear from section 3.3[14].

In order to perform the construction of tensors and spinors as noted above, we have worked out a method for obtaining the induced representations of the little group of rotations, which is invariant under spin subspaces of the Hilbert space. Then, we used this method to obtain concrete expressions for the spin 1/2 Dirac field and for the spin 1 Maxwell field, and we identified possible dynamical equations in this framework.

The spin 1/2 field of the SHP theory (see (3.3)) has the same C, P and T symmetries as the standard Dirac spinor (see (3.17)), as well as the same magnetic dipole moment (see (3.14) and reference[13]). However, the SHP equation is not identical to the Dirac equation. It has the same form as a second order Dirac equation, but the coupling to the spin is purely magnetic through the induced representation. It contains the correct gyromagnetic ratio and, as shown by Bennett[24], accounts as well, to lowest order, for the anomalous moment, and provides the same energy spectrum in the nonrelativistic limit[8].

The spin 1 field that was constructed here can be cast into a quaternionic matrix form. This allows for the identification of the electromagnetic field strength tensor, the gauge condition and the dynamical Lagrangian, as emerging from the anti-symmetric part, the trace and the determinant of $\mathcal{D} \cdot \mathcal{A}$, respectively (see equations (3.25)-(3.27)). This means that the Hamilton principle can be formulated for the electromagnetic theory in terms of a functional of $\mathcal{D} \cdot \mathcal{A}$ (see (3.28)-(3.29)). Although this was carried out explicitly for the Abelian case, this method allows for a natural generalization to non-Abelian gauge groups, thus yielding a matrix functional framework for Yang-Mills type fields.

The vector field and all higher rank fields were shown to be decomposed into definite spin blocks, by the methods outlined in section 2. This fact promotes a viewpoint by which the spin 1/2 representation constitutes not only the fundamental representation, but could also represent a fundamental spin 1/2 particle that all higher spin particles might be built out of. Such a viewpoint is not new, but it emerges entirely from the induced representation formalism.

We have also constructed a Pauli-Lubanski vector which provides a second Casimir operator for the Poincaré group and contains the physical angular momentum of the particle on the orbit of the induced representation. When the average $p^\mu$ in a wave packet is parallel to $n^\mu$, a Lorentz transformation can bring the particle to (approximate)rest in a Stern-Gerlach apparatus which would measure the physical an angular momentum state, consistent with an interpretation given by Aharonov of the $n$-vector[29].

# Appendix

## Wigner, Racah and 3nj coefficients[21]

The 3nj symbols for $n \in \mathbb{N}$ arise from multiplication of Wigner coefficients by one another. We have the following important results:

Wigner coefficients are just the Clebsch-Gordan coefficients referred to in the mathematical literature. These can are related to the 3j symbols according to:

$$\begin{pmatrix} j_1 & j_2 & j \\ \sigma_1 & \sigma_2 & -\sigma \end{pmatrix} = (-1)^{j_1-j_1+\sigma} C^{j_1 j_2 j}_{\sigma_1 \sigma_2 \sigma} \quad (A1)$$

Racah coefficients are defined from the sum of products of four Wigner coefficients. These are related to the 6j symbols, according to:

$$\begin{Bmatrix} j' & k_1 & j'' \\ k_2 & j & k \end{Bmatrix} = (-1)^{j'+k_1+j+k_2} W(j'k_1 j k_2; j''k) \quad (A2)$$

Alternatively, the Racah coefficients can be considered to be the basic elements of the representation theory of $SU(2)$, such that Wigner coefficients may be obtained from them by taking the asymptotic limit of Racah coefficients.

9j and higher order coefficients are defined from the sum of products of three Racah coefficients. Since the Racah coefficients may be obtained from Wigner coefficients, the 9j symbols can also be written as the sum of products of six Wigner coefficients.

All the 3nj symbols for $n > 2$ may be written in terms of composition of Racah (or Wigner) coefficients.

## Recoupling of Angular Momenta[23]

In the process of addition of more than two independent spins, we face the problem of ordering in which the composition takes place. The recoupling of angular momenta has been long connected with the problem of assigning parentheses i.e. to count all possible ways of introducing parentheses into the sum $\vec{j}_1 + \vec{j}_2 + \cdots + \vec{j}_N$ such that each sub-sum is binary.

The binary bracketing of a product of any number $N$ objects is denoted by $(j_1 j_2 \cdots j_N)^B$. It allows for any symbol a unique specification of the intermediate state angular momenta. The number of ways, $a_N$, of introducing parentheses into the product $j_1 j_2 \cdots j_N$ such that each sub-product is binary, is given by the Catalan numbers:

$$a_N = \frac{1}{N}\binom{2N-2}{N-1} \text{ for } N = 2,3,\ldots \text{ (A3)}$$

In including the possibility to permute the spins $j_1 j_2 \cdots j_N$ into $j_{i_1} j_{i_2} \cdots j_{i_N}$, where $i_1 i_2 \cdots i_N$ is a permutation of $1, 2, \ldots, N$, we must replace the Catalan numbers with:

$$c_N = N! \, a_N = (N)_{N-1} \text{ for } N = 2,3,\ldots \text{ (A4)}$$

Binary coupling was shown to be in one-to-one correspondence with binary trees, which are connected acyclic graphs of degree 2. The eigenvalue of the total spin, $j$, defines the root of the tree, the kinematically independent spins of the particles, $j_1 j_2 \cdots j_N$, define terminal points, while the internal points are related to the $N-2$ intermediate-state angular momenta, that will be denoted $k_1 k_2 \cdots k_{N-2}$. These operators are the compensation needed for the total spin basis of $N+2$ ($N$ Casimir operators, $\{\vec{J}_a\}$, and the $\vec{J}^2, J_3$ operators, to describe the full symmetry given by $2N$ Casimirs and their projections. Altogether, the total spin basis defines the complete set of mutually commuting observables of the system:

$$\left\{ \vec{J}_a^{\,2} (a = 1,\ldots,N), \vec{K}_\lambda^{\,2} (\lambda = 1,\ldots,N-2), \vec{J}^2, J_3 \right\} \text{ (A5)}$$

Taking the eigenvalues $k_1 k_2 \cdots k_{N-2}$ to be fixed, we are constrained to count only:

$$d_N = \frac{c_N}{2^{N-1}} = (2N-3)!! \text{ for } N = 2,3,\ldots \text{ (A6)}$$

Thus, we may partition the set of binary bracketings $\{(j_1 j_2 \cdots j_N)^B\}$ into a number of subsets equal to $d_N$, such that each subset contains $2^{N-1}$ elements yielding the state vectors:

$$|(i_1 i_2 \cdots i_N)(j_1 j_2 \cdots j_N)^B (k_1 k_2 \cdots k_{N-2}) j \sigma \rangle \equiv |(i)(j_i)^B (k) j \sigma \rangle \text{ (A7)}$$

Now, each state of the composite system may be written as a linear combination of direct-product states:

$$|(i)(j_i)^B (k) j \sigma \rangle = \sum_{\sum_i \sigma_i = \sigma} C^{(i)}_{(k)} \binom{j_i}{\sigma_i} |j_1, \sigma_1\rangle |j_2, \sigma_2\rangle \cdots |j_N, \sigma_N\rangle \text{ (A8)}$$

with the $C^{(i)}_{(k)}$ being a product of $N-1$ Wigner coefficients for $SU(2)$. Therefore, if one wishes to transform from one scheme to another, one should calculate the transition amplitudes:

$$\langle (i')(j_{i'})^{B'} (k') j' \sigma' | (i)(j_i)^B (k) j \sigma \rangle = \sum_{\sum_i \sigma_i = \sigma} \sum_{\sum_{i'} \sigma_{i'} = \sigma'} C^{(i')}_{(k')} \binom{j_{i'}}{\sigma_{i'}} C^{(i)}_{(k)} \binom{j_i}{\sigma_i} \text{ (A9)}$$

The multiplication of Wigner coefficients by one another, as in Eq. (A9), gives rise to known patterns that are called 3nj symbols for $n \in \mathbb{N}$. The process of addition of $N$ spins gives rise to at most $3(N-1)j$ symbols.